\documentclass{aa}
\usepackage{txfonts}
\usepackage{apj2aa}
\usepackage{float,epsfig,psfig}
\usepackage{pstricks}

\def\vac#1{}

\begin{document}

\title{The Impact of High Pressure Cluster Environment on the X-ray Luminosity of Coma Early-type Galaxies}
\author{A. Finoguenov\inst{1,3} \and F. Miniati\inst{2}}


\institute{Max-Planck-Institut f\"ur extraterrestrische Physik,
             Giessenbachstra\ss e, 85748 Garching, Germany, alexis@mpe.mpg.de
\and
Max-Planck-Institut f\"ur Astrophysik,
Karl-Schwarzschild-Str. 1, 85741, Garching, Germany, fm@mpa-garching.mpg.de
\and
Space Research Institute, Profsoyuznaya 84/32, Moscow, 117810, Russia
} 
\authorrunning{Finoguenov \& Miniati}
\titlerunning{Ellipticals in Coma cluster}
\date{Received  2004, January 29; accepted 2004, March 10}
\abstract{We present an observational study of 
the $L_{\rm X}$ vs $L_{\rm B} \sigma^2$ relation for early-type
galaxies in the Coma cluster based on the XMM-Newton survey data. 
Compared to a similar relation for a sample dominated by field
early-type galaxies, the Coma cluster galaxies show a flatter
slope. Our calculations show that adiabatic compression produces a
flattening in the $L_{\rm X}$ vs $L_{\rm B}\sigma^2$ relation that is in
remarkable 
agreement with the observed effect. Our scenario is further supported by the
observed compactness of the X-ray emission of Coma galaxies.
 }

\maketitle
\section{Introduction}

X-ray emission from early-type galaxies has been subject of investigations
and theoretical modeling ever since its discovery with the Einstein
telescope (Forman, Jones, Tucker 1985). The X-ray emission appears in part
to be diffuse in nature and to originate in the galaxy hot interstellar
medium (ISM); and in part to be due to a population of X-ray point sources,
namely low mass X-ray binaries characterized by a hard spectrum.  In fact,
the X-ray spectra are typically well fit with a two-component model: a soft
$\sim$1 keV thermal component for the hot ISM and and a hard $\sim$1.4
photon index power law contribution for the point sources (Irwin et
al. 2003). The latter component is systematically more important for fainter
objects (Matsushita 2001).

The temperature of the ISM is proportional to the stellar velocity
dispersion and, although with some scatter, the coefficient of
proportionality is close to unity. In fact, the main energy input to the hot
ISM is thought to be due to thermalization of kinetic energy associated with
stellar mass loss, in addition to energy input from explosions of SN Ia
(Mathews 1989; Ciotti et al. 1991; David et al. 1991). However, in some
cases the galaxy ISM is found to be considerably hotter than the stellar
velocity dispersion would indicate, as for example in the case of galaxies
in the center of massive groups (Brown \& Bregman 1998; Matsushita 2001).

In this respect the relation between X-ray and optical properties is
important in order to understand the processes regulating the evolution of
the hot gas. In fact, early on it was realized the existence of a
correlation between the X-ray and the blue band stellar luminosity. Both the
log-slope of the correlation and the significant scatter about it are
physically meaningful and need to be properly understood and modeled.

In addition, X-ray and optical properties of ellipticals are used 
in combination for testing different scenarios of galaxy formation.
Thus, Kodama \& Matsushita (2001) used the continuity of optical properties
across X-ray extended and compact early type galaxies to show that,
if the different X-ray emission properties of the two classes are 
related to the structure of the underlying dark matter potential,
then the formation of stars in these galaxies must predate the epoch
when the potential structure was established. 

In this {\it Letter} we address the impact of high pressure cluster
environment on the diffuse X-ray emission of early type galaxies. For the
purpose we use the recent X-ray catalog of Coma cluster galaxies based on
XMM-Newton survey and presented in Finoguenov et al. (2004a) and compare
with the results reported in Matsushita (2001). In addition, we concentrate
on X-ray compact galaxies as defined in Matsushita (2001). We find that
X-ray emission from early-type galaxies in Coma is systematically higher
when compared to their counterparts in the field or in loose groups, such as
those presented in Matsushita (2001). We show that compression induced by
the high pressure ICM is sufficient to reproduce the observed effect and
point out that such environmental effects should be taken into account when
modeling the X-ray properties of these galaxies. 

This {\it Letter} is structured as follows. In \S\ref{s:data} we describe
the data; in \S\ref{rela:se} we plot $L_{\rm X}$ vs $L_{\rm B}\sigma^2$ for
Coma galaxies illustrating the main result of this letter; in \S\ref{s:th}
we attempt an interpretation of the observational results by estimating the
effect of intracluster gas on X-ray luminosity; we conclude with
\S\ref{s:con}.  We use $H_0=70$~km~s$^{-1}$~Mpc$^{-1}$ throughout.

\section{X-ray data} \label{s:data}

\includegraphics[width=8.5cm]{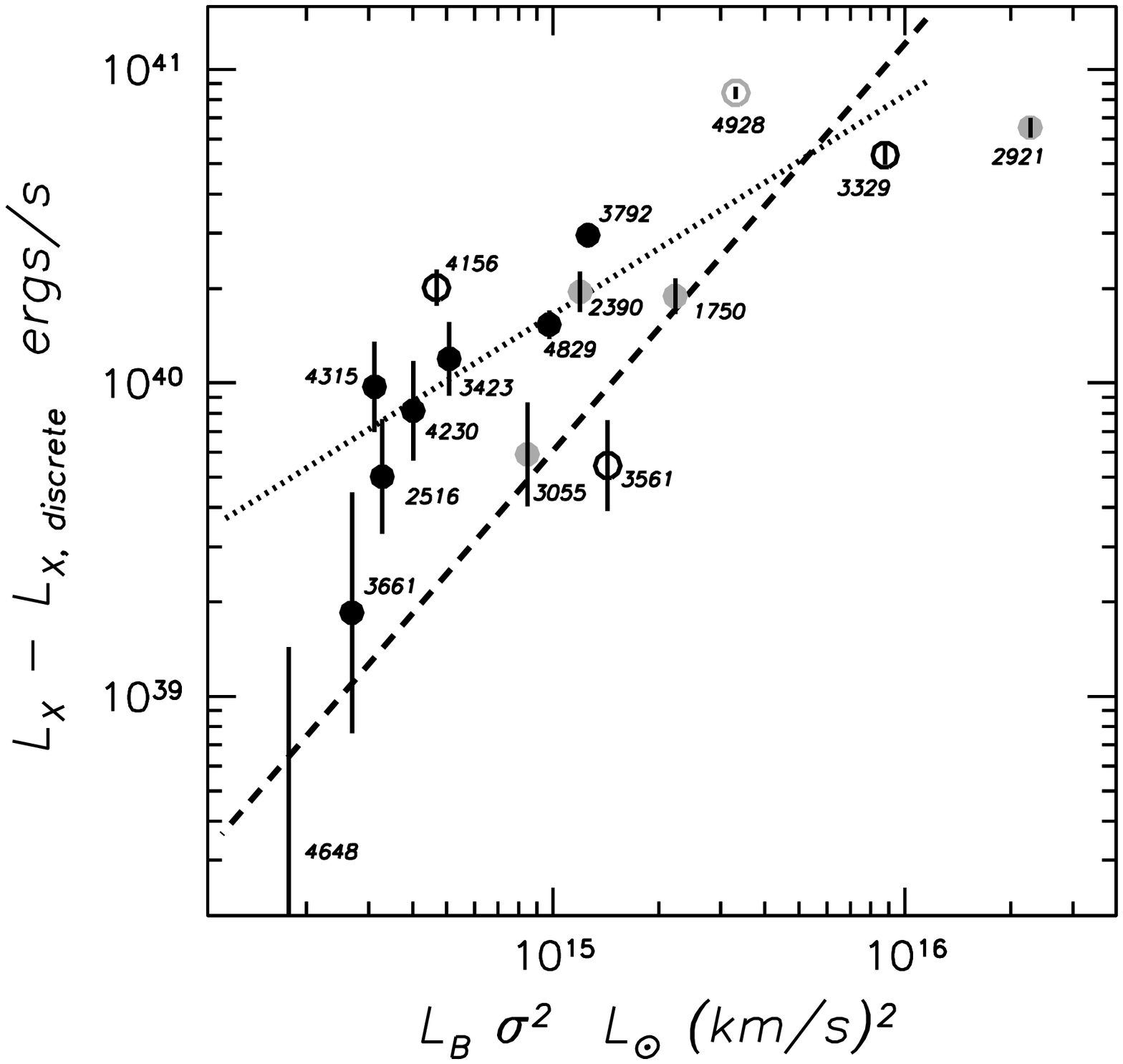}
\figcaption{$L_{\rm X}-L_{\rm B}\sigma^2$, using measurements of $\sigma$
from J{\o}rgensen (1999). Contribution from discrete point sources has been
subtracted (see text for details).  Open symbols denote the sources also
identified as FIRST (radio) sources. Grey points correspond to boxy
ellipticals and black to disky.  Dashed line is a fit to a similar relation
for the field ellipticals (Matsushita 2001) and dotted line is the expected
curve if diffuse X-ray emission balances the energetic input from the
stellar mass loss, that is $\propto \dot{M}_{*}\sigma^2\propto L_{\rm
B}\sigma^2$ (e.g. Ciotti et al. 1991). This curve also represents the
flatter relation expected as a result of the effects of ICM pressure (see \S
\ref{s:th}).  Luminosities are calculated assuming a 100 Mpc distance to the
Coma cluster, appropriate for $H_0=70$~km~s$^{-1}$~Mpc$^{-1}$. Numbers on
the plot show the denumerations of galaxies in the catalog of Godwin et
al. (1983).
\label{f:scsgm}
}

XMM-Newton survey of the Coma cluster (Briel et al. 2001) covered a
region of 1.89 square degrees. In Finoguenov et al. (2004a) over six hundred
sources with extent less then $10^{\prime\prime}$ have been detected, sixty
of which were identified with galaxies that are spectroscopically confirmed
members of the Coma cluster. For the purpose of this paper we have selected a
subsample of early-type galaxies, for which measurements of the velocity
dispersion exists in the literature (J{\o}rgensen 1999). For these galaxies,
we have also found a measurement of the structural parameter, which we
further use to divide the sample into disky and boxy ellipticals (Saglia et
al. 1993; Mehlert et al. 2000). 

Current AGN activity may influence the X-ray luminosity in galaxies
(Finoguenov et al. 2004b), although this should not be the case for
our sample as its members are exclusively characterized by thermal
colors (Finoguenov et al. 2004a). Past AGN activity can still be
revealed by the presence of radio emission. Thus, we have marked with
a special symbol the galaxies in our sample that are also found in the
FIRST survey (Becker et al. 1995).  The sample we will consider in the
following includes the Coma cluster galaxies with the following
denumerations in the catalog of Godwin et al. (1983; GMP): 2390, 2921,
3055, 3329, 3423, 3792, 4156, 4230, 4315, 4829, 4928, 1750, 2516,
3561, 4648, 3661.  For each galaxy we estimate the X-ray luminosity in
the 0.5--2 keV band from the diffuse component. The contribution from
the unresolved point sources has been subtracted using estimates based
on the well resolved (by Chandra) observations of this population in
the nearby elliptical M84 (Finoguenov \& Jones 2001). A correction for
the aperture of XMM-Newton source extraction, was also introduced as
described in Finoguenov et al. (2004a). For most but the two X-ray
faintest sources, subtraction of the unresolved component introduces
no difference in the inferred flux.  In part, this is caused by the
small extent of the diffuse X-ray emission of Coma ellipticals
(Vikhlinin et al. 2001), which results in a thermal to unresolved
point source flux ratio that is higher in the inner regions of these
galaxies. In fact, in Coma galaxies the spatial distribution of the
X-ray emission from LMXB is similar to the stellar light and extends
out to four effective radii, while the diffuse X-ray emission is more
centrally concentrated, typically within an effective radius.

\section{$L_{\rm X}$ vs $L_{\rm B}\sigma^2$ for Coma cluster early-type galaxies} \label{rela:se}

In Fig.\ref{f:scsgm} we plot the diffuse X-ray luminosity versus
L$_{\rm B}\sigma^2$, the product of the stellar blue band luminosity times
its central velocity dispersion. The sample is the one discussed in
the previous section.
The main result of this {\it Letter} is the measurement of a 
significantly flatter slope in the plotted relation with respect 
to the one reported for field galaxies  (Matsushita 2001).

Matsushita (2001) already found a correlation between $L_{\rm X}$ (using a
slightly different energy band, 0.2--2 keV) and $L_{\rm B}\sigma^2$ for a
sample of compact early-type galaxies including 30 field galaxies and 12
cluster galaxies (3 members of Fornax cluster plus 9 members of Virgo
cluster).  Galaxies with emission extending beyond four effective radii were
excluded from her study. As shown in Fig. \ref{f:scsgm} Coma galaxies also
follow a $L_{\rm X}-L_{\rm B}\sigma^2$ relation but with a significantly
different slope. Chandra observations already reported lower X-ray
luminosities for the two central ellipticals (Vikhlinin et al. 2001),
consistent with our plot. In addition, here we show that numerous other
galaxies exhibit {\it higher} X-ray fluxes than expected from the relation
presented in Matsushita (2001), a feature that was not identified in
previous studies.

We note the presence of significant scatter in the plot in Fig.
\ref{f:scsgm}.  In order to investigate its origin, we have inspected the
X-ray images of the sample galaxies, presented in Finoguenov et al. (2004a).
The emission of all the sources is within two effective radii, so they all
fulfill the compactness criterion of Matsushita. However, some of the Coma
galaxies are much more compact than that, e.g. the extent of the emission in
the GMPs 2921 and 3329 is a few percent of the effective radius (Vikhlinin
et al. 2001), which we denote as ultracompact. Other ultracompact X-ray
sources, include GMPs 2921, 3329, 4156, 2390, 4829, 3055, 3423, 4315, 4230,
3792. Their emission appears nearly point-like and limits on the spatial
extent are lower than half the effective radius. With exception of GMP 3055,
which has a factor of two lower X-ray flux and GMP 4156, which has a factor
of two higher X-ray flux, all of them are found on the main relation for
Coma cluster galaxies. GMP 3792 has a secondary point source associated with
another galaxy, which explains its factor of 1.5 higher luminosity. Some
galaxies exhibit extended emission or/and emission strongly displaced from
the center. These are GMPs 2516, 1750, 3561, 3661, 4648. All of them are
found on the relation for field ellipticals. This evidence may suggest that
the galaxies departing from the main relation are experiencing some
disruption such as gas stripping effects. Unlike two bright cluster galaxies
(BCG)in Coma (GMPs 3329 and 2921), BCG of the infalling cluster, GMP 4928
(NGC 4839), has not been separated from the surrounding gas and although it
also exhibits an ultracompact source, we tentatively ascribe it to galaxies
with diffuse emission.

\begin{table}[ht]
\caption{Parameter values for $L_{\rm X} - L_{\rm B} \sigma^2$ relation$^{\dag}$}\label{t:fit}
\begin{tabular}{lcccc}
\hline
sample & A & B & A$^{\flat}$ & B$^{\flat}$  \\
\hline
all      & $1.05_{-0.01}^{+0.01}$ & $1.13_{-0.23}^{+0.29}$ & $1.12_{-0.30}^{+0.36}$ & $1.12_{-0.34}^{+0.40}$\\
--GMP4928& $1.00_{-0.02}^{+0.02}$ & $1.08_{-0.23}^{+0.30}$ & $1.07_{-0.29}^{+0.39}$ & $1.07_{-0.38}^{+0.45}$\\
--radio   & $1.03_{-0.05}^{+0.07}$ & $1.16_{-0.22}^{+0.28}$ &
$1.17_{-0.36}^{+0.64}$ & $1.19_{-0.48}^{+0.70}$\\
radio  & $1.32_{-0.43}^{+0.48}$& $0.93_{-0.38}^{+0.61}$& \multicolumn{2}{c}{unconstrained}\\
disky    & $1.33_{-0.18}^{+0.31}$ & $1.40_{-0.33}^{+0.47}$ & $1.56_{-0.54}^{+2.35}$ & $1.43_{-0.62}^{+0.90}$\\
boxy     & $1.17_{-0.18}^{+0.18}$ & $0.78_{-0.14}^{+0.16}$ & \multicolumn{2}{c}{unconstrained}\\
ultracompact & $1.63_{-0.01}^{+0.01}$ & $0.48_{-0.01}^{+0.01}$ & $1.62_{-0.26}^{+0.24}$ & $0.47_{-0.18}^{+0.14}$\\
--ultracompact & $0.75_{-0.01}^{+0.01}$ & $1.71_{-0.06}^{+0.07}$ &
\multicolumn{2}{c}{unconstrained}\\
\hline
\end{tabular}

\begin{enumerate}
\item[{$^{\dag}$}]{\footnotesize ~ $L_{\rm X}=A \times 10^{40} {\rm ergs} \; s^{-1}
    ({L_B \sigma^2 \over 10^{15} L_\odot {\rm km} \; {\rm s}^{-1}})^B$. Errors are given
    at 68\% confidence level.} 

\item[{$^{\flat}$}]{\footnotesize ~ Bootstrap method is used to estimate the
    mean and the variance.}
\end{enumerate}
\end{table}

In order to quantify the measured effect and possible variations associated
with particular properties of the galaxies in our sample, in Tab.\ref{t:fit}
we present results from a linear regression analysis in $log(L_{\rm
X})-log(L_{\rm B}\sigma^2$) space of the relation $L_{\rm X}=A \times
10^{40}$ ergs s$^{-1}$ $({L_{\rm B} \sigma^2 \over 10^{15} L_\odot {\rm km}
\; {\rm s}^{-1}})^B$. Our method allows for errors in both variables and
intrinsic scatter (Akritas \& Bershady 1996). We use the bisector method to
determine the best-fit parameters (Isobe et al. 1990). The resulting values
are reported in columns (2--3) of Tab.\ref{t:fit}. To estimate the effect of
selection of data points we perform an additional analysis, calculating the
mean and the dispersion in the best-fit values, obtained through 10000
bootstrap realizations. These values are reported in columns (4--5) of
Tab.\ref{t:fit}. The samples we consider, listed in the first column of
Tab.\ref{t:fit}, correspond to: the complete set of galaxies; the same
excluding GMP 4928; the same excluding galaxies with identified radio
counterpart; the set of disky galaxies; the set of boxy galaxies;
ultracompact (located well within half the effective radius) X-ray emission;
diffuse X-ray emission. Comparing the results for the different samples
should reveal how the studied relation changes for, and is affected by,
galaxies with different properties. Removing GMP 4928 has no effect on our
results which are determined by the bulk of the studied objects.  Similarly,
results are not significantly affected by galaxies with identified radio
counterpart. Contrary to their position on the $L_{\rm X}-L_{\rm B}$
relation (Bender et al. 1989), disky galaxies are not underluminous in the
sense of $L_{\rm X}-L_{\rm B}\sigma^2$, in agreement with Kodama \&
Matsushita (2000). However, disky and boxy galaxies yield somewhat different
slope, with disky galaxies exhibiting a steeper relation. However, bootstrap
realizations indicate that this result could be reproduced by selection of
systems. Galaxies with ultracompact emission exhibit slope flatter than a
relation for the field ellipticals in Matsushita (2001) at the highest
significance.

In the following, we will analytically derive the influence of the
external pressure on the observed slope of the $L_{\rm X}-L_{\rm
B}\sigma^2$ relation.

\section{Impact of intracluster pressure}\label{s:th}

The X-ray emitting gas in elliptical galaxies is expected to be in
conditions of hydrostatic equilibrium so that a given gas pressure arises in
response to, and to balance out, the pull of the gravitational force.

However, within the environment created by a much more massive structure,
such as a rich cluster of galaxies, the pressure can be quite higher that
required by the conditions of hydrostatic equilibrium. Therefore, the galaxy
interstellar gas can be further compressed. A direct confirmation of the
pressure equilibrium between the ISM of two Coma ellipticals and the ICM is
presented in Vikhlinin et al. (2001). If the compression is adiabatic,
entropy is conserved and as a result of a pressure increment, $\Delta P \ge
0$, the following thermodynamic variations are expected to take place
\begin{eqnarray}
P & \longrightarrow & P(1+\Delta P/P) \\
\rho & \longrightarrow & \rho (1+\Delta P/P)^{1/\gamma} \\
T & \longrightarrow & T (1+\Delta P/P)^{(\gamma-1)/\gamma}
\end{eqnarray}
where $P, \rho, T$ are pressure, density and temperature respectively,
$\gamma$ is the gas adiabatic index, $\Delta P=P_{\rm ICM}-P$, and 
$P_{\rm ICM}$ is the external pressure which in this specific case 
corresponds to the intracluster pressure of Coma cluster.

If the X-ray luminosity were simply due to thermal bremsstrahlung emission,
then 
\begin{eqnarray}
L_{\rm X} \propto \rho T^{1/2} M_{\rm gas}
\longrightarrow L_{\rm X} (1+\Delta P/P)^{(\gamma+1)/2\gamma}.
\end{eqnarray}
where $M$ is the galactic interstellar gas mass.  However, X-ray emission in
the 0.5--2 keV band of XMM-Newton EPIC pn as well as 0.2--2 keV band of
ROSAT PSPC detectors is dominated by L-shell iron emission lines. It turns
out that due to changes in the ionization equilibria, for metallicities of
order $0.1-0.5$ solar, a temperature increase does not affect the emitted
X-ray power in the selected energy band. More precisely, the flux slightly
increases for metallicities $\sim 0.1$ but decreases for metallicities $\ge
0.5$. The ISM metallicities of all these galaxies are not precisely known,
but are expected to be in this range (Finoguenov \& Jones 2000, 2001), as
measured for two brightest Coma ellipticals (Vikhlinin et al. 2001). So for
the current calculation we will assume that the flux is unchanged under
temperature variations. Thus, the changes in the X-ray luminosity induced by
compression are simply those due to the density enhancements to the first
power
\begin{equation} \label{lxt.eq}
L_{\rm X} \longrightarrow L_{\rm X} (1+\Delta P/P)^{1/\gamma} 
\propto  L_{\rm X} T^{-1/(\gamma-1)}
\end{equation}
where in the last passage we have used the $P-T$ relation for an adiabatic
gas.

In order to infer the change in the $L_{\rm X} - L_{\rm B}\sigma^2$ relation we need to
somehow express the quantity $L_{\rm B}\sigma^2$ in terms of some thermodynamic
variable, like the temperature for example. According to scaling relations
for virialized systems, the velocity dispersion should be of order of the
virial temperature $T$. In addition we expect that the B-band luminosity be
proportional to the stellar mass loss, that feeds the X-ray luminosity of
ellipticals (Mathews 1989); if the latter is a fixed fraction (proportional
to) the halo mass, then invoking again scaling relations we obtain
$L_{\rm B}\propto T^{3/2}$ and $L_{\rm B}\sigma^2\propto T^{5/2}$.

Combining this result with Eq. (\ref{lxt.eq}), we find
\begin{equation} \label{lxt.eq2}
L_{\rm X} = (L_{\rm B}\sigma^2)^\alpha \longrightarrow 
(L_{\rm B}\sigma^2)^{\alpha-{2 \over 5(\gamma-1)}}.
\end{equation}
where $\alpha$ is the slope in absence of ICM compression effects, like the
case measured by Matsushita (2001). For an adiabatic index $\gamma=5/3$ the
slope of the above relation is expected to flatten by an amount $3/5$. The
flattened curve as predicted by this calculations is plotted in
Fig. \ref{f:scsgm} as a dotted line and given the simplicity of our model
the match to the observed effect is quite good.

\section{Discussion and conclusions}\label{s:con}

In this paper we have studied a sample of early type galaxies from Coma
cluster. We have shown that the diffuse X-ray emission of such galaxies when
plotted against the blue band stellar luminosity times the stellar central
velocity dispersion, follows a flatter curve than similar galaxies in the
field do (Matsushita 2001). Further, we have shown that, starting from the
$L_{\rm x}$ vs $L_{\rm B} \sigma^2$ relation measured by Matsushita (2001),
we can recover the flatter relation for the galaxies in Coma cluster by
accounting for the effects of adiabatic compression from the hot ICM.

$L_{\rm B} \sigma^2$ represent the maximum energy input associated with
stellar mass loss (Matsushita 2001, Ciotti et al. 1991).  According to our
calculations the line \mbox{$L_{\rm X} =3.2\times 10^{40} \; {\rm ergs}\;
{\rm s}^{-1}\; \alpha_{\rm x} \;{ L_{\rm B} \sigma^2 \over 10^{15}\; L_\odot
\; ({\rm km} \; {\rm s}^{-1})^2}$} in the $L_{\rm X}-L_{\rm B}$ diagram,
where $\alpha_{\rm x} \simeq 0.4$ is the ratio of 0.5--2 keV X-ray to
bolometric emissivity, corresponds to the maximum X-ray emissivity if indeed
stellar mass loss is the energy source that feeds the diffuse X-ray
emission. In fact, this line roughly draws an upper envelope for the compact
sources in Matsushita's (2001) sample. Interestingly, Coma galaxies reported
in this letter tend to lay on such curve. This is consistent with the above
idea about the source of energy for the X-ray emission. In this respect, it
is worth pointing out that numerous Coma galaxies in our sample can be
characterized as having a ``ultracompact'' X-ray structure and, therefore,
according to Matsushita, should be underluminous. Despite this, they also
clump near the same theoretical curve based on the assumption that the
heating of the galaxies ISM is supplied by the stellar mass loss.

It is worth pointing out that in deriving the X-ray scaling relations
for the early-type galaxies, we have not included the contribution of SN Ia
to gas heating (Ciotti et al. 1991).  However, according to recent stellar
age determinations, field early-types have a systematically younger age than
the cluster population (e.g. Terlevich, Forbes 2002), which implies a
stronger SN Ia activity there (Greggio \& Renzini 1983). Thus, if SN Ia
played a dominant role in determining the X-ray luminosity of early-types,
we would expect cluster galaxies to exhibit lower luminosities, which is not
observed.

There are theoretical arguments and observational evidence supporting the
idea that ram pressure stripping is one of the principal mechanisms
affecting the morphology of X-ray galaxies (e.g. Vollmer et al. 2001,
Bravo-Alfaro et al. 2001, Finoguenov et al. 2004a; see also Finoguenov et
al. 2004c).  Gas stripping, however, implies a reduction of gas content of
the cluster galaxies which, in absence of other effects, leads to a decrease
of the level of X-ray emission. We have demonstrated that in fact X-ray
emission in Coma cluster galaxies is on average higher than in their field
counterparts with similar stellar properties. This is sensibly explained by
ICM compression as outlined in the previous Section, which, therefore, adds
to the wide variety of processes that affect the evolution of X-ray
galaxies.

Finally, as Coma galaxies are more luminous than those in the field because
of the high outer pressure, some galaxies in the field may likewise be
underluminous, because they expanded when the energy input from mass loss is
sufficiently high in comparison to the depth of their potential well.  It
will be interesting to further investigate this possibility in the future
especially in connection with the new developments in the study of low-sigma
early-type galaxies (Burkert \& Naab 2003).

\begin{acknowledgements}

The paper is based on observations obtained with XMM-Newton, an ESA science
mission with instruments and contributions directly funded by ESA Member
States and the USA (NASA). The XMM-Newton project is supported by the
Bundesministerium f\"{u}r Bildung und Forschung/Deutsches Zentrum f\"{u}r
Luft- und Raumfahrt (BMBF/DLR), the Max-Planck-Gesellschaft (MPG) and the
Heidenhain-Stiftung, and also by PPARC, CEA, CNES, and ASI. The authors
thank Luca Ciotti, Bernd Vollmer, Kyoko Matsushita, and an anonymous referee
for enlightening discussions and comments, resulted in a substantial
improvement of this work. AF acknowledges support from BMBF/DLR under grant
50 OR 0207 and MPG. FM work was partially supported by the Research and
Training Network `The Physics of the Intergalactic Medium', EU contract
HPRN-CT2000-00126 RG29185.
\end{acknowledgements}

\bibliographystyle{aabib99}

\end{document}